\documentclass[aps,prd,twocolumn,nofootinbib,floatfix,unsortedaddress]{revtex4}
\DeclareSymbolFont{matha}{OML}{txmi}{m}{it}
\DeclareMathSymbol{\varv}{\mathord}{matha}{118}
\usepackage{setspace}
\usepackage{graphicx}
\usepackage{dcolumn}
\usepackage{multirow}
\usepackage{stackrel}
\usepackage{subfigure}
\usepackage{times,mathptm}
\usepackage{float}
\usepackage{color}
\usepackage{amsmath,amsfonts}
\usepackage{mathptmx}
\usepackage{mathrsfs}
\usepackage{bbm}
\usepackage{bm}
\usepackage{xfrac}

\newcommand{\beq}{\begin{equation}}
\newcommand{\eeq}{\end{equation}} 
\newcommand{\bea}{\begin{eqnarray}}
\newcommand{\eea}{\end{eqnarray}}

\def\lsim{\mathrel{\rlap{\lower4pt\hbox{\hskip1pt$\sim$}}
    \raise1pt\hbox{$<$}}}
\def\gsim{\mathrel{\rlap{\lower4pt\hbox{\hskip1pt$\sim$}}
    \raise1pt\hbox{$>$}}}
\newcommand{\adot}{\dot{a}}
\newcommand{\pdot}{\partial_t \vph}

\renewcommand{\H}{\mathcal{H}}

\newcommand{\cS}{\mathcal{C}}

\newcommand{\cF}{\mathcal{F}}

\renewcommand{\d}{\delta}

\renewcommand{\l}{\lambda}

\newcommand{\p}{\phi}
\renewcommand{\b}{\beta}
\renewcommand{\a}{\alpha}

\renewcommand{\o}{\omega}

\newcommand{\vx}{{\vec{x}}}
\newcommand{\vy}{{\vec{y}}}

\newcommand{\vk}{{\vec{k}}}

\newcommand{\n}{\nu}
\newcommand{\m}{\mu}

\newcommand{\g}{\gamma}
\renewcommand{\r}{\rho}
\newcommand{\e}{\epsilon}
\newcommand{\s}{\sigma}
\renewcommand{\k}{\kappa}

\newcommand{\G}{{\cal G}}

\newcommand{\vp}{\vec{p}}

\newcommand{\N}{{\cal N}}

\newcommand{\M}{{\cal M}}
\renewcommand{\th}{\theta}
\newcommand{\rg}{\sqrt{g}}

\newcommand{\vph}{\varphi}
\newcommand{\oh}{\frac{1}{2}}
\newcommand{\oq}{\frac{1}{4}}

\newcommand{\dg}{\dagger}
\newcommand{\non}{\nonumber}
\renewcommand{\t}{\tau}
\newcommand{\rf}[1]{(\ref{#1})}
\newcommand{\ra}{\rightarrow}
\newcommand{\pa}{\partial}
\renewcommand{\vec}[1]{\bm #1}

\usepackage{ulem}

\begin{document}

\title{Inflation and Cyclotron Motion} 

\bigskip
\bigskip

\author{Jeff Greensite}

\bigskip

\affiliation{ \vspace{5pt} Physics and Astronomy Department, San Francisco State
University,   \\ San Francisco, CA~94132, USA}
 
\date{\today}
\vspace{60pt}
\begin{abstract}

\singlespacing

    We consider, in the context of a braneworld cosmology, the motion of the universe coupled to
a four-form gauge field, with constant field strength, defined in higher dimensions.  It is found, under rather general initial
conditions, that in this situation there is a period of exponential inflation combined with cyclotron motion in the inflaton field space.   The main effect of the cyclotron motion is that slow roll conditions on the  
inflaton potential, which are typically necessary for exponential inflation, can be evaded. There are Landau levels associated with the four-form gauge field, and these correspond to quantum excitations of the inflaton field satisfying unconventional dispersion relations.

\end{abstract}

%
%
%
\maketitle

\singlespacing
\section{\label{intro}Introduction}

   Braneworld cosmology is a concept that exists in many variations.  There are versions in which the higher dimensions are compactified, as in the Arkani-Hamed, Dimopoulous, Dvali proposal \cite{ArkaniHamed:1998nn}, or large but warped, as in the Randall-Sundrum model  \cite{Randall:1999vf} and string-motivated DBI inflation \cite{Alishahiha:2004eh,HenryTye:2006uv}.  There is also the intriguing Dvali-Gabadadze-Porrati (DGP) version where the extra dimension is large but nearly flat \cite{Dvali:2000hr}.  Consideration of the four-dimensional effective theory in the DGP model has led to a very general class of four-dimensional galileon models \cite{Nicolis:2008in}
with powers of derivative terms greater than two, for which there now exists an extensive literature (see, e.g., \cite{Burrage:2010cu,Trodden:2011xh,Deffayet:2009mn,Neveu:2016gxp} and references therein).
   
   In this article I would like to describe some interesting features of the following action, describing a brane with standard model particle content evolving in a flat higher-dimensional background, with a coupling of the brane to an external four-form gauge field in the bulk:
\bea
    S &=& {1\over 16\pi G}\int d^4x \sqrt{-g} R  + S_{SM}
\non \\
& & - \int d^4x \sqrt{-g} \Bigl( \oh g^{\m\n} \pa_\m \vph^s \pa_\n \vph^s + V(\vph) \Bigr)
\non \\
& & + {q_0\over 4!} \int d^4x ~ A_{abcd}[\p(x)] \e^{\a\b\g\d} \pa_\a \p^a \pa_\b \p^b \pa_\g \p^c \pa_\d \p^d \; , \non \\
\label{S0}
\eea
where $S_{SM}$ is the action of standard model (and possibly beyond-standard-model) fields, and
\bea
          g_{\m\n} &=& \pa_\m \p^A \eta_{AB} \pa_\n \p^B  ~~~,~~~ A,B = 0,1,...,D
\label{induce}
\eea
is the induced metric of a three brane in a $D+1$-dimensional Minkowski space.  
We adopt the convention that upper case Latin indices run from 0 to $D$,
indices $r,s$ run from $D+1$ to $D+N$, and all other lower case Latin indices run from $0$ to $D+N$.
We also define
\beq
    \p^s = {1\over \s^2} \vph^s \; ,
\eeq
where $\s$ is a constant with dimensions of mass.  The $\vph^s$ fields are a set of $N$ inflaton fields, with $V(\vph)$ the inflaton potential, and $A_{abcd}$ is a potential which is totally antisymmetric in the indices.  It can be thought of as a four-form gauge field in $D+1+N$ dimensions.  The induced metric corresponds to $D+1$ dimensions, however.   

    The main novelty of this formulation is the interaction of the braneworld with an external four-form gauge field in the bulk, and it is the
purpose of this article to describe some possible consequences in an inflationary scenario.  Like the DGP model there are large flat extra dimensions, but unlike that model there is no Einstein-Hilbert action in the bulk.  Unlike Galileon models in general there is no galilean invariance, and the external four-form gauge field singles out special directions in the bulk.  Inflation, in the scenario suggested below, is driven by inflaton fields with an ordinary $V(\vph)$ potential in the inflaton action, rather than by galileon fields. 

   Without the external gauge field, a model with an Einstein-Hilbert action and other fields on the brane seems to have been first considered long ago by Regge and Teitelboim \cite{Regge}.  The first question to ask of a model of this type is whether the equations of motion are equivalent, at the classical level, to standard general relativity at $A_{abcd}=0$. The answer is: not quite.  Denote
 \bea
        E^{\m \n} &\equiv& { \d S[A=0] \over  \d g_{\m\n} }  \non \\
                        &=& \oh \sqrt{-g} \left\{ -{1\over 8\pi G } G^{\m \n}  + T^{\m \n}   \right\} \; .
\eea
Where $T^{\m\n}$ is the stress-energy tensor of the standard model and inflaton fields. Then the field equations resulting from variation of the $\p^A$ at $A_{abcd}=0$ are
\beq
  \eta_{AB} \pa_\m (E^{\m \n} \pa_\n \p^B) = 0 \; .
\label{eom}
\eeq
These equations are obviously satisfied by the Einstein field equations $E^{\m\n} = 0$.  Moreover,  any solution of
$E^{\m\n}=0$ can be embedded locally in a ten-dimensional flat Minkowski space,
although globally an embedding may require still higher dimensions \cite{Clarke}.  But of course there may be also be
solutions of \rf{eom} which are not solutions of the Einstein equations.  A simple (and intriguing) example is pure gravity with a cosmological constant, in which case 
\beq
E^{\m \n} = \oh \sqrt{-g} \left\{ -{1\over 8\pi G } G^{\m \n}  - \l g^{\m\n} \right\}  \; .
\eeq
In this case the equations of motion are certainly solved by de Sitter space, for which $E^{\m\n}=0$.  But flat Minkowski space
is also a solution:  just choose $\p^\m = x^\m,~ \m=0-3$ and $\p^{A>3} =$ constant.  Then $g_{\m\n}=\eta_{\m\n}$,
$G_{\m\n}=0$, and the equations of motion boil down to $\Box \p^A = 0$, which is satisfied trivially.

   A criticism of Deser et at.\ \cite{Deser:1976qt} is that the embedding of a four-manifold is not unique.  Some embeddings of a four-manifold may satisfy the equations of motion \rf{eom}, and some may not.  This fact does not necessarily rule out the embedding formulation of general relativity on experimental grounds; it could simply be that the $E^{\m \n} = 0$ alternative is selected by initial conditions on the $\p^a$.

   When the four-form gauge field is included, there will in general be some deviation from the standard Einstein field equations.  The equations of motion in this case are 
\bea
& &2 \eta_{AB} \pa_\m (E^{\m \n} \pa_\n \p^B)   \non \\
& & \qquad - {q_0\over 4!} F_{Aabcd} \e^{\a\b\g\d} \pa_\a \p^a \pa_\b \p^b \pa_\g \p^c \pa_\d \p^d = 0 \; ,
\label{eom1}
\eea
and
\bea
& & \pa_\m(\sqrt{-g} g^{\m\n} \pa_\n \vph^s) - \sqrt{-g} {\pa V \over \pa \vph^s} \non \\
& & \qquad + {q_0 \over 4! \s^2} F_{sabcd}\e^{\a\b\g\d} \pa_\a \p^a \pa_\b \p^b \pa_\g \p^c \pa_\d \p^d = 0 \; ,
\label{eom2}
\eea
where $F_{fabcd}$ is the field strength
\bea
F_{fabcd} = {\pa A_{abcd} \over \pa \p^f} -  {\pa A_{fbcd} \over \pa \p^a}  +  {\pa A_{facd} \over \pa \p^b} 
-  {\pa A_{fabd} \over \pa \p^c}  +  {\pa A_{fabc} \over \pa \p^d}  \non \\
\label{F}
\eea
corresponding to the four-form gauge field. These are supplemented by the usual equations of motions of the
standard model fields.

In this article I would like to explore the cosmological consequences
of these equations of motion in the simplest non-trivial case, namely, a constant field strength $F_{fabcd}$ in a homogenous isotropic spacetime.  For this purpose it will be sufficient to work in a five-dimensional embedding space, ${A=0,..,4}$, with two inflaton fields $\vph^{5,6}$, and ignoring, at the classical level, all standard model fields.

\section{Inflation}

   It is well known that a four dimensional manifold described by a Friedman-Lemaitre metric can be embedded in five-dimensional space, and for simplicity we adopt the version with zero spatial curvature.  We take the embedding to be
\cite{Rosen, LachiezeRey:2000my}
\bea
         \p^0 &=& \oh \left\{ a(t) + \int^t {dt' \over da/dt'} + a(t) r^2 \right\} \non \\
         \p^1 &=& a(t) r \cos(\th)   \non \\
         \p^2 &=& a(t) r \sin(\th) \cos(\chi)   \non \\
         \p^3 &=& a(t) r \sin(\th) \sin(\chi)   \non \\
         \p^4 &=& \oh \left\{ a(t) - \int^t {dt' \over da/dt'} - a(t) r^2 \right\} \; ,
\label{embed}
\eea
and it is not hard to see that
\bea
   ds^2 &=& \eta_{AB} d\p^A d\p^B  ~~,~~ A,B = 0,1,2,3,4  \non \\
             &=& -dt^2 + a^2(t) (dr^2 + r^2(d\th^2 + \sin^2(\th) d\chi^2))
\eea
is the Friedman-Lemaitre metric.  But let us also suppose that there is a four-form gauge field dependent on the
coordinates $\phi^a$, whose non-zero components are
\bea
        A_{5123}[\p] &=& - \oh B \p^6 \; , \non \\
        A_{6123}[\p] &=& \oh B  \p^5 \; .
\label{B}
\eea
The four-form gauge field $A_{abcd}$ is antisymmetric under permutations of indices, but apart from \rf{B} and components obtained from \rf{B} by permutation, it is assumed that all other components vanish.  This choice leads to a constant non-zero field strength $F_{56123}=B$, and we are interested in exploring the consequences for early-universe dynamics in a situation of this kind.  In this context we also assume the simplest possible inflaton potential
\beq
V[\p] = \oh m^2  \vph^s \vph^s \ .
\label{Vinf}
\eeq

     We begin with the usual simplifying assumptions of spatial homogeneity and isotropy, taking in particular
\beq
           \p^{5,6}(x,y,z,t) = \p^{5,6}(t)    \; ,
\label{isotropic}
\eeq
and $\p^a = 0$ for $a>6$.  In conjunction with \rf{B}, this has the consequence that
\beq
        F_{Aabcd} \e^{\a\b\g\d} \pa_\a \p^a \pa_\b \p^b \pa_\g \p^c \pa_\d \p^d = 0 \; .
\eeq
This is because two of the indices $abcd$ must be 5 and 6, so the expression necessarily includes at least
one space derivative of $\vph^s$, which vanishes according to \rf{isotropic}.
Then the equation of motion $\rf{eom1}$ is satisfied by $E^{\m\n}=0$, which are the standard Einstein field
equations.   For a Friedman-Lemaitre metric, disregarding the other standard model fields, the Einstein equations are 
just the conventional expressions for the $a(t)$ scale factor coupled to a pair of scalar fields: 
\bea
            {\adot^2 \over a^2} &=&  {8\pi G \over 3} \left( \oh \pdot^s \pdot^s + \oh m^2 \vph^s \vph^s \right)  \; , \non \\
            {\ddot{a} \over a}  &=&   {8\pi G \over 3} \left( - \pdot^s \pdot^s + \oh m^2 \vph^s \vph^s \right) \; .
\label{Einstein}
\eea
The equations of motion for the $\vph^s$, however, involve the field strength
\bea
         \pa^2_t \vph^5 -  qB \pdot^6 + 3 {\adot \over a}\pdot^5 + m^2 \vph^5 &=& 0 \; , \non \\ 
         \pa^2_t \vph^6 +  qB \pdot^5 + 3 {\adot \over a}\pdot^6 + m^2 \vph^6 &=& 0 \; ,
\label{eom3}
\eea
where $q = q_0/\s^4$. It is not hard to verify consistency of \rf{Einstein} and \rf{eom3}.
 
   If we set $\adot/a = 0$ and $m^2=0$ in \rf{eom3}, then these equations are obviously the equations of motion of a charged particle moving, in the $\vph^5-\vph^6$ plane, under the influence of a magnetic field orthogonal to
that plane; i.e.\ this is cyclotron motion.  If we instead set $qB = 0$, then these are the equations used in simple models of inflation. In models of that type it is normally important to impose slow roll conditions, which imply either a large initial value for the inflaton field, or else, unlike \rf{Vinf}, a very flat potential (see, e.g.,
Chapter 8 in \cite{Peter:1208401}).  For the simple potential \rf{Vinf} these slow roll conditions boil down to
\beq
           \vph^s \vph^s \gg {1 \over 6 \pi G} \; ,
\label{sroll}
\eeq
i.e.\ a large initial field.

\begin{figure}[t!]
\subfigure[]  
{   
 \label{xy}
 \includegraphics[scale=0.5]{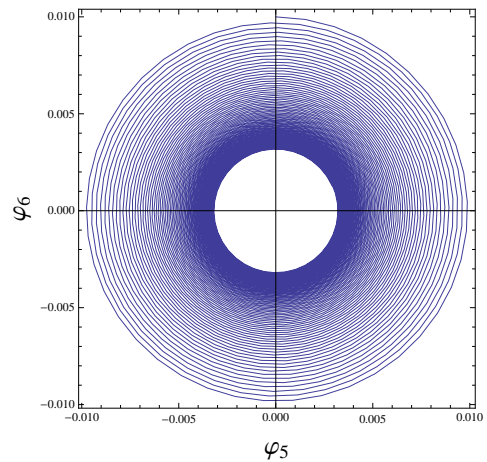}
}
\subfigure[]  
{   
 \label{adot}
 \includegraphics[scale=0.5]{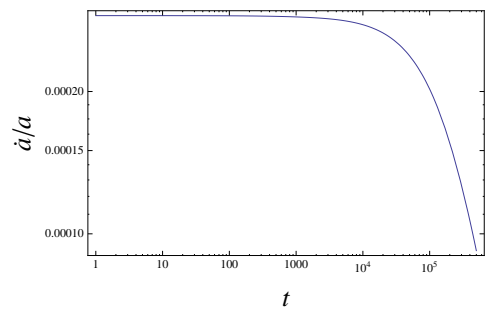}
}
\subfigure[]  
{ 
 \label{addot}
 \includegraphics[scale=0.5]{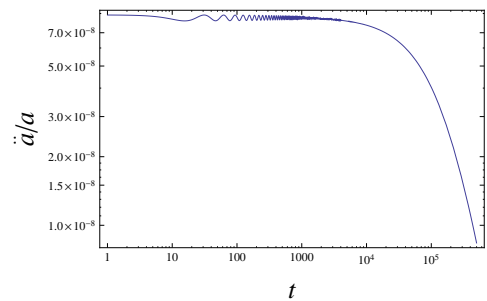}
}
\caption{Numerical solution of the evolution equations \rf{Einstein} and \rf{eom3}, with parameters and initial conditions
\rf{starting}.  (a) trajectory in the $\vph^5-\vph^6$ plane; 
(b) log-log plot of $\dot{a}/a$ vs.\ time $t$; (c) log-log plot $\ddot{a}/a$ vs.\ time $t$.  Note that the log-log
plots of $\dot{a}/a$ and $\ddot{a}/a$ vs.\ time $t$ are almost flat in the period $1<t<10^4$, indicating a period of
exponential expansion, in this case with about 100 e-foldings.}
\label{long}
\end{figure}

   The model we are discussing has a fairly large space of parameters  and initial conditions $\{qB, m^2,\vph^s(0), \pa_t \vph^s(0)\}$ but the time development is typically a spiral in the $\vph^5-\vph^6$ plane.  What may be of interest is the fact that for $qB \ne 0$ it is possible to have a period of approximately exponential inflation, with a large number of e-foldings, even when the slow-roll condition \rf{sroll} is strongly violated.\footnote{It should be noted, however, that there are other mechanisms for easing the slow roll conditions in the context of a braneworld cosmology, cf.\ \cite{Maartens:1999hf}.}  A single example should suffice.  Working in Planck units, we choose parameters and initial conditions
\bea
   qB &=& 0.2 ~~,~~ m^2 = 2\times 10^{-4} \; , \non \\
  \vph^5(0)&=& 0 ~~~,~~~ \vph^6(0) = 10^{-2}  \; , \non \\
 (\pa_t \vph^5)_{t=0} &=& 0   ~~~,~~~ (\pa_t \vph^6)_{t=0} = 0 \;  .
\label{starting}
\eea       
The resulting spiral evolution in the $\vph^5-\vph^6$ plane is shown in Fig.\ \ref{xy}, with $\dot{a}/a$ and $\ddot{a}/a$ vs.\
cosmic time $t$ shown in Figs.\ \ref{adot} and \ref{addot} respectively.  The expansion is very nearly a simple exponential up to
$t \approx 10^4$ in Planckian units, which is evident in the rather flat curves on the log-log plots, and the fact that
\beq
{\ddot{a} \over a} \approx \left({\dot{a}\over a }\right)^2
\eeq
in this period.  Expansion continues after this period, however,  resulting in a total of about 100 e-foldings by $t=10^6$.

\begin{figure}[htb]   
 \includegraphics[scale=0.5]{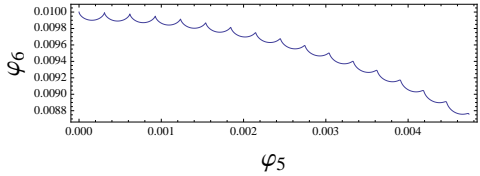}
\caption{The trajectory of Fig.\ \ref{xy} at the beginning of the time evolution, in period $0<t<500$, showing
the effect of the ``Lorentz force,'' directed away from the center of the $\vph^5-\vph^6$ plane.}
\label{xy_init}
\end{figure}

\begin{figure}[htb]   
 \includegraphics[scale=0.5]{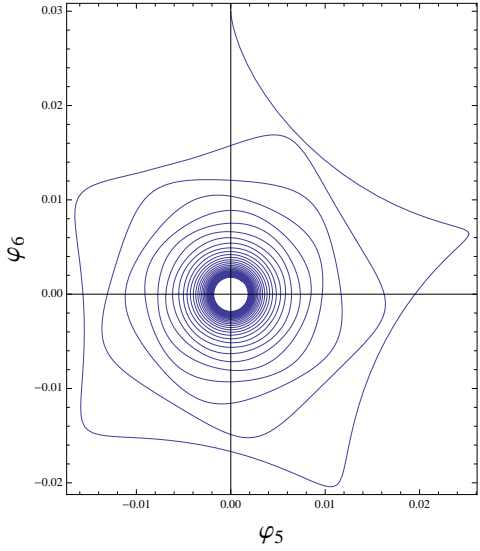}
\caption{Trajectory in the $\vph^5-\vph^6$ plane for parameters ${qB=-1,m^2=0.5,\p^6(0)=0.03}$.}
\label{xy1}
\end{figure}


The potential $V(\p)$ is responsible for a force towards the origin of the $\vph^5-\vph^6$ plane, while the
``Lorentz force'' due to the four form gauge field is directed away from the origin.  Eventually these forces balance to produce a
circular motion, spiraling towards the center.  To see this, we plot the initial stage of the evolution in Fig.\ \ref{xy_init}.
In the absence of the gauge field, the system simply falls to the center, oscillating around the $\vph^6$ 
axis, and, because slow roll conditions are not satisfied, there is no inflationary period.  The Lorentz force, however, deflects the initial fall to the center into an arc, and this interplay between the central potential, the Lorentz force, and gravitational friction continues until the inward and outward forces sum to a centripetal force for (roughly) circular motion, with gravitational friction causing a gradual spiral to the origin.  The trajectory resulting from a quite different set of
parameters is shown in Fig.\ \ref{xy1}.  While this last example does not lead to many e-foldings, it does very clearly display the initial interplay of forces, leading to an eventual spiral towards the origin.  


\section{\label{LL}Landau Levels}

    After inflation, the constant field strength of the four-form gauge field still has an effect at the quantum level, in the
form of Landau excitation levels of the quantized $\vph$ fields.  We will see that these excitations satisfy a rather unusual dispersion relation.

    We consider the post-inflationary period at some time $t_0$ where $\adot/a$ is negligible, $a(t)\approx R$.
With $\p^A$ given by the embedding \rf{embed}, and $A_{abcd}$ as in \rf{B},  we have 
\bea
\lefteqn{{q_0\over 4!} A_{abcd}[\p(x)] \e^{\a\b\g\d} \pa_\a \p^a \pa_\b \p^b \pa_\g \p^c \pa_\d \p^d} & &
\non \\
&=& q A_{s123} \e^{0ijk} \pa_t \vph^s \pa_i \p^1 \pa_j \p^2 \pa_k \p^3 
\non \\
&=& q A_s \pa_t \vph^s  (R^3 r^2 \sin\th) \; ,
\eea
where $A_s \equiv \s^2 A_{s123}$.  The factor of $R$ can be absorbed into a coordinate redefinition, and we then consider
quantizing the action
\bea
S_\vph &=& \int d^4x  \Bigl( \oh \pdot^s \pdot^s - \oh \nabla \vph^s \cdot \nabla \vph^s 
\non \\
& & \qquad  - \oh m^2 \vph^s \vph^s  + q A_s(\vph) \pdot^s  \Bigr) ,
\label{Svph}
\eea
where again the index $s=5,6$.  The corresponding Hamiltonian is
\bea
H &=& \oh \int d^3x  \left\{ \left(  p_s - qA_s\right)^2 + (\nabla \vph^s)^2 + m^2 \vph^s \vph^s \right\} \; ,
\eea
and $\vph^s,p_{s'}$ have standard quantization conditions.  Define
\bea
    \o_k &=& \sqrt{k^2 +\oq q^2 B^2 + m^2} \non \\
    \vph^s(x) &=& \int {d^3 k \over (2\pi)^3} {1 \over \sqrt{2  \o_k}} (a_s(k)e^{i\vk \cdot \vx}  + a^\dg_s(k)e^{-i\vk \cdot \vx}) \non \\
    p_s(x) &=&   \int {d^3 k \over (2\pi)^3} \sqrt{2  \o_k} {1\over 2i}(a_s(k)e^{i\vk \cdot \vx}  - a^\dg_s(k)e^{-i\vk \cdot \vx}) \; ,
\label{omega}
\eea
with the usual commutation relations
\beq
[a_s(k_1),a^\dg_r(k_2)]=(2\pi)^3 \d^3(\vk_1-\vk_2) \d_{rs}
\eeq  
Then
\bea
H &=& \int {d^3 k \over (2\pi)^3} \Bigl\{ \o_k (a^\dg_s(k) a_s(k) + \d^3(0) )   \non \\
    & & + i \oh qB(a^\dg_5(k) a_6(k) - a^\dg_6(k) a_5(k)) \Bigr\} \; .
\eea
Introduce
\bea
      b_1(k) &=& {1\over \sqrt{2}}\bigg( a_5(k) + i a_6(k) \bigg) \non \\
      b_2(k) &=& {1\over \sqrt{2}}\bigg( a_5(k) - i a_6(k) \bigg) \; .
\eea
which again have the usual commutation relations
\beq
[b_i(k_1),b_j^\dg(k_2)]=(2\pi)^3 \d^3(\vk_1-\vk_2) \d_{ij}
\eeq 
with indices $i,j=1,2$.  The Hamiltonian takes the form
\bea
H &=& \int d^3k \left\{ \o_k (b^\dg_i(k) b_i(k) + \d^3(0) ) \right. \non \\
    & &\left.  + \oh qB(b^\dg_1(k) b_1(k) - b^\dg_2(k) b_2(k)) \right\} \; ,
\eea
and the corresponding spectrum is
\bea
E &=& \sum_k \left\{ \sqrt{k^2 +\oq q^2 B^2 + m^2} \bigg(n_1(k) + n_2(k)\bigg) \right. \non \\ 
   & & \left. + \oh qB\bigg(n_1(k) - n_2(k)\bigg) \right\} + E_0 \; ,
\label{spectrum}
\eea
where $n_1(k),n_2(k)$ are occupation numbers, $E_0$ is the ground state energy, and the sum runs over
momenta with non-zero occupation numbers.  We also find, by standard manipulations,
the conserved total momentum
\beq
    P_i = \sum_k k_i \Bigl(n_1(k)+n_2(k) \Bigr) \; .
\eeq

   Were it not for the term proportional to $qB$ in \rf{spectrum}, the spectrum would simply consist of two types of particles of mass 
\beq
 M'  = \sqrt{\oq q^2 B^2 + m^2} \; .
\label{mprime}
\eeq
Instead, defining $M=\oh qB$, it is seen that excitations which are eigenstates of both $H$ and $P_i$ (with momentum eigenvalues $k_i$)  satisfy dispersion relations
\bea
      E_1(k) &=& \sqrt{k^2 + M^2 + m^2} + M\; , ~~~\mbox{and} \non \\
      E_2(k) &=& \sqrt{k^2 + M^2 + m^2} - M \; ,
\eea
respectively, which is clearly at odds with the relativistic expression for a free particle.  But of course these excitations are not
free particles, and the Lagrangian \rf{Svph} they derive from is not Lorentz invariant, or even (unlike Newtonian mechanics) boost invariant.  It is the external four-form gauge field which singles out a preferred time direction (much as, e.g., an ordinary background magnetic field along the $z$-axis would introduce a preferred spatial direction for objects sensitive to that field), and the only remaining space-time symmetries are rotation and time/space translation invariance. Therefore the breaking of both Lorentz and boost invariance, so far as these inflaton excitations are concerned, is not a surprise.  The question is how this breaking might manifest itself.

\section{Properties of Landau level excitations}

\subsection{Group velocity}

    To begin with, consider how a wavepacket corresponding to a single ``heavy'' Landau excitation of energy $E_1(k)$, or a ``light'' Landau
excitation of energy $E_2(k)$, and momentum $\vk$, will propagate in time.  Let $|\vk,j\rangle$ correspond to a particle eigenstate
of energy and momentum $E_j(k), \vk$ respectively, with conventional normalization
\bea
            |\vk,j\rangle &=& \sqrt{2\o_k} b_j(\vk) |0 \rangle  \non \\
            |\vx,j\rangle &=& \int {d^3k \over (2\pi)^3} e^{-i\vk \cdot \vx} |\vk,j \rangle \ ,
\eea
and we consider initial wavepackets of the form
\bea
        |\psi_j \rangle_{t=0} &=& \int {d^3k \over (2\pi)^3} {1 \over \sqrt{2\o_k}|} f(k) |\vk,j \rangle \non \\
        \psi_j(\vx,t=0) &=& \langle \vx,j | \psi_j \rangle_{t=0} \non \\
                               &=& \int  {d^3k \over (2\pi)^3} f(k) e^{i\vk \cdot \vx} \ .
\eea 
Then at a later time
\bea
       \psi_j(\vx,t) &=&    \langle \vx,j | e^{-iHt} | \psi_j \rangle_{t=0} \non \\       
                         &=&    e^{-i (3-2j)M t}  \int  {d^3k \over (2\pi)^3} f(k) e^{i(\vk \cdot \vx - \o_k t)}  \ .
\eea
From this we conclude that wavepackets of both heavy and light Landau excitations (we might as well call them ``landons") propagate with a group velocity $\varv = p /\o_p$  appropriate to a particle of mass $M' \approx M$ (for $m\ll M$).  On the other hand, at low momenta in the frame singled out by the external four-form gauge field, 
\bea
           E_1(k) &\approx& {k^2 \over 2 M} + 2M + {m^2 \over 2M} \non \\
           E_2(k) &\approx& {k^2 \over 2 M} +  {m^2 \over 2M} \ ,
\label{E12nr}
\eea
which means that the rest energy of the heavy landons is approximately $2M$, while that of the light landons
is approximately $m^2/2M$.  

\subsection{Scattering in a gravitational field}

Because of the mismatch between the inertial
mass in the momentum-dependent $k^2/2M$ term and the rest energy, we may expect an apparent violation of the principle of
equivalence, if it would be possible to somehow observe the motion of these excitations in a gravitational field.  This can be verified by calculating the differential scattering cross section of heavy and light landons in the weak gravitational field of a static massive object of mass $\M$.   

   Let $g_{\m \n} = \eta_{\m \n} + h_{\m \n}$ with
\bea
           g_{00} &=& -\left( 1 - {2G\M\over r}\right) ~~,~~ g_{ii} = \left( 1 + {2G\M\over r}\right)  \non \\
           g_{\m \n} &=& 0 ~~~ (\m \ne \n) \ ,
\eea
be the metric corresponding to the massive object at the origin, at distances $r$ such that the gravitational field is weak.  For our purposes it is sufficient to ignore this restriction on $r$, unless we are interested in large angle scattering.   We first need the interaction Hamiltonian
to lowest order in $G\M$.  For this we consider the part of the total action $S' = S_\vph + S_A$ containing $\vph$, where
\bea
   S_\vph &=& - \int d^4 \sqrt{-g} (\oh g^{\m \n} \pa_\m \vph^s \pa_\n \vph^s + \oh m^2 \vph^s \vph^s)  \non \\
   S_A &=& \int d^4x ~ qA_s \e^{0ijk} \pa_t \vph^s \pa_i \p^{1} \pa_j \p^{2}\pa_k \p^{3}  \ .
\label{SvphA}
\eea
Expanding $S_\vph$ to first order in $G\M$ we have
\bea
S_\vph &=& \int d^4x  \left\{ \oh\left( 1 + {4G\M\over r}\right) \pa_t \vph^s \vph^s - \oh (\nabla\vph^s)\cdot (\nabla\vph^s) \right. \non \\
             & &  \left.  - \oh\left( 1 + {2G\M\over r}\right) m^2 \vph^s \vph^s \right\} \ .
\eea
To compute $S_A$ to leading order in $h_{\m \n}$ we use
\beq
        S_A \approx S_A(h=0) + \int d^4x {\d S_A \over \d g_{\m\n}} h_{\m \n} \ .
\eeq
Now $S_A$ depends on the metric through the $\pa_\m \phi^A$.  As noted already, there is no unique mapping from the metric to 
the three-brane coordinates, but this turns out not to be a problem.  Choose any mapping $g_{\m\n} \ra \pa_\m \phi^A$ and observe that,
acting on any functional of the metric,
\bea
          {\d \over \d (\pa_\m \phi^A)} &=& {\pa g_{\a\b} \over \pa(\pa_\m \phi^A)} {\d \over \d g_{\a \b} } \non \\
                                                    &=&  2 \eta_{A B} \pa_\a \phi^B {\d \over \d g_{\a \m}} \ ,
\eea
which can be inverted to give
\beq
          {\d \over \d g_{\m \n}} = \oh g^{\m \a} \pa_\a \phi^A {\d \over \d(\pa_\n \phi^A) } \ .
\eeq
Applying this operator to $S_A$ in \rf{SvphA}, we find
\bea
   \d S_{A} &=& \int d^4x \left({\d S_A \over \d g_{\m\n}}\right)_{g_{\a \b}=\eta_{\a \b}} h_{\m \n} \non \\
                 &=& {3\over 2} \int d^4x ~ {2G\M \over r} q A_s \pa_t \vph^s \ .
\eea
Altogether
\bea
S' &=& \int d^4x  \left\{ \oh\left( 1 + {4G\M\over r}\right) \pa_t \vph^s \pa_t \vph^s - \oh (\nabla\vph^s)\cdot (\nabla\vph^s) \right. \non \\
             & &  \left.  - \oh\left( 1 + {2G\M\over r}\right) m^2 \vph^s \vph^s  + \left( 1 + {3G\M\over r}\right)q A_s \pa_t \vph^s  \right\} \ .
             \non \\
\eea
We go to the Hamiltonian formulation, introducing canonical momenta conjugate to the $\vph^s$
\beq
           p_s = \left( 1 + {4G\M\over r}\right) \pa_t \vph^s + \left( 1 + {3G\M\over r}\right)q A_s \pa_t \vph^s \ ,
\eeq
leading to a Hamiltonian operator
\begin{widetext}
\bea
H &=& \int d^3x \left\{ \oh  \left( 1 + {4G\M\over r}\right)^{-1} \left(p_s - \left( 1 + {3G\M\over r}\right)q A_s\right) 
\left(p_s - \left( 1 + {3G\M\over r}\right)q A_s\right) \right. \non \\ & &
 \left.   + \oh (\nabla\vph^s)\cdot (\nabla\vph^s) +  \oh\left( 1 + {2G\M\over r}\right) m^2 \vph^s \vph^s \right\}  
 \non \\
 &=& H_0 + \int d^3x \left\{ - {2G\M \over r} (p_s - gA_s)(p_s - gA_s) + {G\M \over r} m^2 \vph^s \vph^s 
 - {3G\M \over r} q A_s (p_s - gA_s) \right\} \ .
\eea
Then the Hamiltonian density in the interaction picture, to first order in $G\M$, is \footnote{Note that in the interaction picture 
the $G\M=0$ operator identification $p_s = \pa_t \phi_s + q A_s$ must be used for the interaction Hamiltonian density.}
\beq
   \H_I = -{2G\M \over r} \bigg\{ \pa_t \vph^s \pa_t \vph^s - \oh m^2 \vph^s \vph^s 
                    + {3\over 2} M(\vph^5 \pa_t \vph^6 - \vph^6 \pa_t \vph^5) \bigg\} \ .
\eeq
Using interaction picture operators
\bea
        \vph^5(x) &=& \int {d^3 k \over (2\pi)^3} {1 \over \sqrt{2  \o_k}} 
                               {1\over \sqrt{2}}  \bigg( b_1(k)e^{i(\vk \cdot \vx-E_1(k)t)}  + b^\dg_1(k)e^{-i(\vk \cdot \vx-E_1(k)t)} 
                        + b_2(k)e^{i(\vk \cdot \vx-E_2(k)t)}  + b^\dg_2(k)e^{-i(\vk \cdot \vx-E_2(k)t)} \bigg)
 \non \\
        \vph^6(x) &=& \int {d^3 k \over (2\pi)^3} {1 \over \sqrt{2  \o_k}} 
                               {1\over \sqrt{2} i}  \bigg( b_1(k)e^{i(\vk \cdot \vx-E_1(k)t)}  - b^\dg_1(k)e^{-i(\vk \cdot \vx-E_1(k)t)} 
                        - b_2(k)e^{i(\vk \cdot \vx-E_2(k)t)}  + b_2^\dg(k)e^{-i(\vk \cdot \vx-E_2(k)t)} \bigg)      \ ,                                
\eea
\end{widetext}
we can compute matrix elements
\beq
     \langle \vp_2, j | \int d^4x {\cal H}_I | \vp_1, j \rangle \ ,
\eeq
and from there it is a standard exercise to calculate the differential cross sections for the
heavy/light (${j=1,2}$) Landau excitations in the specified gravitational field.   The answer is 
\beq
            \left({d\s \over d\Omega}\right)^{grav}_{type~j} = (G\M)^2 {(E_j^2(p) - \oh m^2 \mp {3\over 2} 
            M E_j(p))^2 \over p^4 \sin^4(\th/2)} \ ,
\label{cross_section12}
\eeq
where the minus sign is for type 1 and the plus sign for type 2 landons.  The type-changing cross sections, in which an initial type 1 landon scatters into a type 2 final state or vice versa, both vanish.  We note that for normal scalar fields, i.e.\
${E_p = \sqrt{p^2+m^2}, ~ A_s=M=0}$, eq.\ \rf{cross_section12} agrees with the gravitational cross section previously obtained
by Golowich et al.\ \cite{Golowich:1990gp}.  

   Now let us go to the low-momentum $p^2 \ll m^2 \ll M^2$ limit.  For comparison, the differential cross section for a particle of 
mass $m$ in a potential
\beq
         V(r) = -{\l \over r} \ ,
\eeq
computed via the Born approximation in non-relativistic quantum mechanics is the familiar Rutherford result
\beq
     \left({d\s \over d\Omega}\right)^{Ruth} = {1\over 4} \l^2 {1 \over m^2 v^4 \sin^4(\th/2)} \ .
\eeq
For normal scalar particles ($\oh q B = M = 0$), using \rf{cross_section12} with the approximations \rf{E12nr} in the low momentum limit, the gravitional cross section can be expressed
\beq
     \left({d\s \over d\Omega}\right)_{normal}^{grav} = {1\over 4} (G\M m)^2 {1 \over m^2 v^4 \sin^4(\th/2)} \ ,
\eeq
which, comparing to the Rutherford potential, corresponds to scattering from the potential
\beq
            V(r) = - {G\M m \over r} \ .
\eeq
In other words, the gravitational mass and the inertial mass are the same.  In contrast, for landons of types 1 and 2, 
eq.\ \rf{cross_section12} becomes in the limit $p^2 \ll m^2 \ll M^2$
\bea
     \left({d\s \over d\Omega}\right)_{type~1}^{grav} &=& {1\over 4} (G\M 2M)^2 {1 \over M^2 v^4 \sin^4(\th/2)} \non \\
    \left({d\s \over d\Omega}\right)_{type~2}^{grav} &=& {1\over 4} \left(G\M {m^2\over 2M}\right)^2 {1 \over M^2 v^4 \sin^4(\th/2)}   \ . 
\eea
This is a result that we might have guessed.  By comparison to the Rutherford cross-section, the gravitational masses of both 
types 1 and 2 landons are equal to their rest energies, which (for $m\ll M$) are $2M$ and $m^2/2M$ respectively, while the inertial mass, in accordance with its appearance in group velocity, is approximately $M$ in both cases.

    The principle of equivalence, of course, asserts the identity of gravitational
and inertial mass, which would seem to be badly violated for both heavy and light landons.  Indeed,  in the present scenario, if it were possible to drop a heavy and a light landon from the top of a tall building and observe how they propagate, the heavy landon would accelerate at $2g$, while the light landon would drift downwards
(assuming $m \ll M$) only very slowly, with acceleration $(m^2/2M^2) g$. These odd effects should be viewed as only an apparent violation of the equivalence principle, arising due to interaction with an external four-form gauge field that singles out a particular time direction.  A rough analogy might be the retardation in the gravitational acceleration of a falling conducting ring in the presence of a constant magnetic field directed parallel to gravitational field.  If we were unaware of the external field, this might also seem like a violation of the principle of equivalence, rather than a manifestation of Lenz's Law.  In the present situation, the external four-form gauge field makes a contribution to the landon rest energies which cannot be absorbed into the
inertial masses, resulting in both an unusual dispersion relation, and a seeming violation of the equivalence principle.

\subsection{Energy density in the early Universe}

   If the inflaton field couples only to gravity and the external four-form gauge field, as assumed from the beginning in \rf{S0}, then
observations of the sort just mentioned would be difficult carry out, and it may be more useful to look for signatures of the unconventional dispersion relations in the early universe, due to an unconventional equation of state.   Since it requires an energy of at least $4M$ to pair-create the heavy excitations, and assuming $M$ is $O(1)$ in Planck units, then after inflation the number density of these objects is fixed.  Assuming a dilute ideal gas, the equation of state is conventional:
\beq
          \rho = n\left(2M + {m^2 \over 2M} \right) + {3\over 2} P  \ ,
\eeq
where $\rho, n$ are energy and number density, respectively, and $P$ is pressure. 
The result follows from Boltzmann statistics, plus the fact that, in a non-relativistic regime where \rf{E12nr} applies, momentum degrees of freedom enter quadratically.  Hence the equipartition theorem applies, and the result is no different than that of a monatomic ideal gas, with particles of rest energy $2M + m^2/2M$.   Heavy excitations would contribute to deceleration in the matter-dominated era, but their contribution cannot be easily distinguished from that of other types of matter.

    The situation is more interesting with respect to light excitations.  It is assumed that the rest energy $m^2/M$ is so small that the number of these excitations is not fixed in the hot environment of the early universe\footnote{At least, the number is not fixed if there are any interaction terms in the inflaton potential.  If this is not the case and the number is fixed, then the analysis is the same as for an ideal
gas with particle rest mass $m^2\over 2M$.  Taking $m^2/M \ll P$, result is $\rho \approx {3\over 2} P$, which, it will be seen, is the same as the grand canonical result derived below.} and the chemical potential can be taken to be zero.
In that situation, as with photons, it is necessary to carry out the analysis in a grand canonical ensemble.  Following the usual analysis, the logarithm of the grand canonical partition function ${\cal Z}$ is
\bea
         \log {\cal Z} &=& -V \int {d^3 k \over (2\pi)^3} \ln\left(1 - e^{-\b E_2(k)} \right) \non \\
                        &=& \b V P   \ ,
\label{logZ}
\eea
with $E_2(k)$ defined in \rf{E12nr}.  The energy density is
\beq
           \r = \int  {d^3 k \over (2\pi)^3} {E_2(k) \over e^{\b E_2(k)} - 1}   \ .
\eeq
We assume that in the early universe $m \ll k \ll M$, and observe that
\bea
{d\over dk} \ln\left(1 - e^{-\b E_2(k)}\right) &=& {\b \over e^{\b E_2(k)} -1} {d\over dk}\left( {k^2 +m^2 \over 2M} \right) \non \\
&=& {\b k \over M} { 1 \over e^{\b E_2(k)} -1}
\eea
Applying this identity we have
\bea
       \r &=& {4 \pi \over (2\pi)^3} \int_0^\infty dk ~ k^2 E_2(k) {M \over \b k}{d\over dk} \ln\left(1 - e^{-\b E_2(k)}\right) \non \\
          &=& {4 \pi \over (2\pi)^3}  {M \over \b} k E_2(k) \ln\left(1 - e^{-\b E_2(k)}\right) \Big|_0^\infty \non \\
           & &         -  {4 \pi \over (2\pi)^3}  {M \over \b}\int_0^\infty dk \left({d \over dk}k E_2(k)\right) \ln\left(1 - e^{-\b E_2(k)}\right) \non \\
\eea
The boundary terms go to zero linearly with $k$ as $k\ra 0$, and exponentially to zero like $\exp(-\b k^2/2M)$ as $k \ra \infty$.  Carrying out the derivative inside the integral we have
\bea
  \r &=& - {4 \pi \over (2\pi)^3} {3\over 2\b} \int_0^\infty dk ~ k^2  \ln\left(1 - e^{-\b E_2(k)}\right)  \non \\ 
     & &   - {4 \pi \over (2\pi)^3} {1\over 2\b} \int_0^\infty dk ~ m^2  \ln\left(1 - e^{-\b E_2(k)}\right)
\eea
The magnitude of the integrand of the second integral only exceeds the magnitude of the integrand of the first integral 
for $k < m/\sqrt{3}$.   However, the logarithm is $O(1)$ up $k \approx \sqrt{2M/\b}$, after which it falls exponentially.  
Therefore, if ${m^2/2M \ll 1/\b}$, then the interval $m/\sqrt{3} < k <  \sqrt{2M/\b}$ is far larger than the interval $0<k<m/\sqrt{3}$. The second integral is therefore negligible compared to the first, and, comparing to \rf{logZ}, we have
\bea
\r &=&  -  {3\over 2\b} \int {d^3k \over (2\pi)^3} \ln\left(1 - e^{-\b E_2(k)}\right)  \non \\
   &=& {3\over 2} P
\eea

  An equation of state with $P=w\r$ leads, in an FRW metric, to a dependence $\r \sim a^{-3(1+w)}$.  In our case, with $w={2\over 3}$, that implies $\r \sim a^{-5}$. This raises the interesting possibility, since ordinary radiation energy density goes as $a^{-4}$, that following inflation there might have been a ``Landau level-dominated'' era, just prior to the radiation-dominated era.   Of course, to pin down the time of transition between these two eras it would be necessary to know an additional cosmological parameter $\Omega_{Landau}$ in the Friedmann equation, and at the moment this number is unknown.  It is understood that since the light landons only manifest their effects through gravitation, they could only be in thermal equilibrium with other particles when gravity is relatively strong, i.e.\ near the Planck time.

\subsection{Causality}

On a flat $g_{\m\n}=\eta_{\m\n}$ background, the field commutators are
\bea
[\vph^5(x),\vph^5(y)]   &=& \cos(M(x_0-y_0)) \{D_{M'}(x-y) - D_{M'}(y-x)\} \non \\
{[\vph^6(x),\vph^6(y)]} &=& \cos(M(x_0-y_0)) \{D_{M'}(x-y) - D_{M'}(y-x)\} \non \\
{[\vph^5(x),\vph^6(y)]} &=& \sin(M(x_0-y_0)) \{D_{M'}(x-y) - D_{M'}(y-x)\} \ , \non \\
\label{comm}
\eea
where
\beq
     D_{M'}(x-y) = \int {d^3k \over (2\pi)^3} {1 \over 2\o_k} e^{i(\vk \cdot (\vx-\vy) - \o_k (x_0-y_0))} 
\eeq
and $\o_k, M'$ were defined in \rf{omega} and \rf{mprime} respectively.
For spacelike separations $x-y$, the difference  
\beq
\Delta D_{M'} = D_{M'}(x-y) - D_{M'}(y-x)
\eeq
vanishes, and hence
the field commutators vanish, consistent with causality.  It has been assumed that the $\vph^s$ fields are only observable via
their coupling to gravity; i.e.\ through the stress-energy tensor.  The commutation relations \rf{comm} also imply that
spacelike separated stress-energy operators commute.

\section{Conclusions}

   It has been shown that, within a braneworld scenario in which the three-brane is coupled to a four-form gauge field, a cosmological version of cyclotron motion can result in a period of exponential inflation with an appropriate number of e-foldings, even in the absence of
the usual slow-roll conditions on the inflaton potential.  The mechanism is that the tendency of the inflaton field to fall to the minimum of the potential is countered by a Lorentz force in the inflaton field space. We also find a spectrum of quantum excitations of the inflaton fields, essentially a cosmological version of Landau levels, satisfying unusual dispersion relations.   One consequence of the unconventional dispersion relations is the possible existence of a Landau level-dominated era, with energy density $\r \sim a^{-5}$, preceding the radiation-dominated era. 

    So far only the simplest aspects of this scenario have been discussed.  The fluctuation spectrum, production of standard model particles, and possible observational signatures in the CMB, call for further investigation.


\appendix*
\section{No-Brane Version}

   We may also consider the action \rf{S0} without the assumption of an embedding \rf{induce} and corresponding braneworld cosmology.  In other words, the $\p^a$ are simply taken to be ordinary scalar fields, which may have a potential of some kind, and are degrees of freedom completely distinct from the metric, which is fundamental rather than induced.  While this alternative setup may not be so relevant to inflationary cosmology, the formulation may still be interesting as a generalization of the Lorentz force law to Wheeler-DeWitt superspace. 
   
   The action of a charged spinless point particle in interaction with an electromagnetic field is
\bea
      S &=& -m\int d\t \sqrt{-g_{\m \n} {dx^\m \over d\t}{dx^\n \over d\t}} + q \int dx^\m A_\m  \; ,
\label{ppaction}
\eea
leading to the equation of motion
\bea
    & &  g_{\m\n} {d^2 x^\n \over ds^2} + \oh \left( {\pa g_{\m \a} \over \pa x^\b}
           + {\pa g_{\m \b} \over \pa x^\a}  - {\pa g_{\a \b} \over \pa x^\m} \right)
      {dx^\a \over ds}{dx^\b \over ds} = {q\over m} F_\m \; , \non \\
& & \qquad \qquad \mbox{where} ~~~
  F_\m = \cF_{\m \n} {dx^\n \over ds} \; ,
\label{pointparticle}
\eea
and $\cF_{\m \n}$ is the electromagnetic field strength tensor.  This is simply the Lorentz force law in curved spacetime. 

     We restrict the discussion to purely bosonic fields, including gravity.  To fix notation,  let $\{q^A(\vx),p_A(\vx),~A=1,2,...,n_f\}$ denote the canonical conjugate variables  with the non-gravitational fields scaled by an appropriate
power of Newton's constant so as to be dimensionless.  The index $A$ now runs over all spatial indices and quantum numbers carried by the fields. In the absence of the four form field $A_{abcd}$, the first-order ADM action has the form
\bea
      S_{ADM} &=& \int d^4x \; [p_A \pa_t q^A - N\H_x - N_i\H^i_x] \; ,
\non \\
      \H_x &=& \k^2 G^{AB} p_A p_B + \rg U(q) \; ,
\non \\
      \H^i_x &=& O^{iA}[q,\pa_x] p_A \; ,
\label{ADM}
\eea
and the dynamics is given by Hamilton's equations plus the constraints $\H_x=\H^i_x=0$.  In the case of pure gravity,
the correspondence with standard notation is
\bea
        \{A=1-6\} &\leftrightarrow& \{(i,j),~i \le j \}
\non \\
             q^A(x) &\leftrightarrow& g_{ij}(x)
\non \\
             p_A(x) &\leftrightarrow&
\left\{ \begin{array}{rr}
             p^{ij}(x)~~~~~(i=j) \\
           2 p^{ij}(x)~~~~~(i<j) \\
        \end{array} \right.
\non \\
             G_{AB}(x) &\leftrightarrow&  G^{ijnm}(x)
\non \\
          \rg U &=& - {1 \over \k^2}\rg {~}^{(3)}R
\non \\
            \H^i &=& -2p^{ik}_{~~;k} \; ,
\eea
where $\rg$ is the determinant of the three-metric $g_{ij}$, $\k^2=16\pi G$, ${~}^{(3)}R$ is the three-dimensional scalar curvature, $G_{ijkl}$ is the DeWitt superspace metric
\beq
   G_{ijkl} = {1 \over 2\rg}(g_{ik}g_{jl}+g_{il}g_{jk}-g_{ij}g_{kl}) \; ,
\eeq
and of course Hamilton's equations plus constraints are equivalent to the Einstein field equations for pure gravity.

   Now let $q^A(x) = \p^A(x)$ for indices $A \in \cS$, where $\cS$ is a subset of indices.  We will denote indices
in this subset by lower-case Latin letters, and the $\p^a$ are a set of scalar fields.  Adding the term
\beq
{q_0\over 4!} \int d^4x ~ A_{abcd}[\p(x)] \e^{\a\b\g\d} \pa_\a \p^a \pa_\b \p^b \pa_\g \p^c \pa_\d \p^d
\label{extra}
\eeq
to the action, and going over to the Hamilitonian formulation, it is readily verified that the expressions for $\H_x,\H^i_x$
are changed by minimal substitution
\beq
           p_a(x) \ra p_a(x) - q_0 A_a(x) \; ,
\eeq
where
\beq
            A_a(x) \equiv {1 \over 3!} A_{abcd}\e^{0ijk} \pa_i \p^b \pa_j \p^c \pa_k \p^d \; .
\eeq
In order that the constraint algebra is satisfied, it is necessary that terms involving $\d A_a / \d \p^f$, which now arise
in the usual Poisson brackets among the $\H_x, \H^i_x$, all cancel.  With a little more effort, those cancellations can also be verified. 
 
     In the case of a standard action containing only bosonic fields, i.e\ including the metric tensor but not the four-form
$A_{abcd}$ field,  it has been shown \cite{Greensite:1995ep} that the geodesic equation derived from the following 
action
\bea
   S_q &=& - \int d\t \; \sqrt{-\G_{(A \vx)(B \vy)}
           {dq^{(A \vx)} \over d \t}  {dq^{(B \vy)} \over d \t} } 
\non \\
&=& - M \int ds \; ,
\label{proper}
\eea
(reminiscent in some ways of the Baierlein-Sharp-Wheeler action \cite{Baierlein:1962zz}) is equivalent to the equations of motion of the standard action in a certain gauge.  In other words, bosonic field equations in general relativity can be
expressed as the geodesic motion of a point particle in Wheeler-DeWitt superspace with a non-standard 
supermetric $\G$.  This is of course in close analogy to Jacobi's principle in mechanics.  The notation is as follows: We define a mixed discrete/continuous index $(A x)$ as a ``coordinate index'' in superspace
\bea
     q^{(A x)}  = \left\{ \begin{array}{cl}
                     \N(x)   & A=0 \\
                     q^A(x)  & A \ne 0 \end{array} \right. \; ,
\eea
with summation convention
\beq
       V_{..(A x)..} W^{..(A x)..} \equiv \sum_{A=0}^{n_f} \int d^3x \;
             V_{..(A x)..} W^{..(A x)..} \; ,
\eeq
and the non-standard supermetric is taken to be
\bea
  \G_{(A x)(B y)} &=&  
          \left[\int d^3x' \; \N \rg U \right]
          {1 \over 4\N(x) \k^2} G_{AB}(x) \d^3(x-y) \; , \non \\
\label{metric}
\eea
while $\G_{(A x)(B y)} =0$ for $A=0$ and/or $B=0$.  With these definitions, it is found \cite{Greensite:1995ep} that the equations of motion which follow from \rf{proper} are the same as those for the standard action in a shift gauge $N_i=0$, with lapse function 
\beq
           N = M { \N \over \int d^3x \N \sqrt{g} U(q) } \; ,
\label{lapse}
\eeq
and $M$ is any constant with dimensions of mass.  The choice of $M$ is essentially a 
choice of affine parameter.

   Adding \rf{extra} to \rf{proper}, the equations of motion are
\begin{widetext}
\beq
     \G_{(A x)(B y)} {d^2 q^{(B y)} \over d s^2} + \oh \left(
     {\d \G_{(A x)(B y)} \over \d q^{(C z)} } +
     {\d \G_{(A x)(C z)} \over \d q^{(B y)} } -
     {\d \G_{(B y)(C z)} \over \d q^{(A x)} } \right)
     {dq^{(B y)} \over ds}{dq^{(C z)} \over ds}
     =  q_0 F_{(A x)} \; ,
\label{geo}
\eeq
\end{widetext}
where $F_{(Ax)}=0$ for indices $A \notin \cS$, while for $A=f \in \cS$
\beq
         F_{(fx)} = {1\over 3!} F_{fabcd}[\p(x)] \e^{ijk0} \pa_i \p^a \pa_j \p^b \pa_k \p^c {\pa \p^d \over \pa s} \; ,
\label{Ff}
\eeq
and $F_{fabcd}$ is given in \rf{F}.  Inserting the supermetric \rf{metric} in \rf{geo}, one finds that these are the equations of motion that follow from the standard action \rf{S0} (excluding fermionic fields) in the shift gauge $N_i=0$ and lapse function \rf{lapse}.

   Equations \rf{geo} and \rf{Ff} are the suggested extension of the Lorentz force law to Wheeler-DeWitt superspace,
reducing to the usual bosonic field equations (including gravity) for $A_{abcd}=0$.  Of course these equations of motion are
no different from those obtained from the action \rf{S0}, only dispensing with \rf{induce} and treating the metric components as fundamental degrees of freedom.  It should be noted, however, that the $\p^a$ fields in this no-brane formulation have no particular correlation with coordinates in a Friedman-Lemaitre metric, and for this reason we do not expect the kind of cyclotron motion and inflation that is seen in the braneworld version.

\bigskip

\acknowledgments{I would like to thank Kristan Jensen for helpful discussions.  This research is supported by the U.S.\ Department of Energy under Grant No.\ DE-SC0013682. }

\bibliography{bff}

\end{document}